\documentclass[
 reprint,
 superscriptaddress,
preprintnumbers,
 amsmath,amssymb,
 aps,
 prl
]{revtex4-2}

\usepackage{gensymb}
\usepackage{graphicx}
\usepackage{dcolumn}
\usepackage{bm}
\usepackage[hidelinks]{hyperref}
\usepackage[mathlines]{lineno}
\usepackage{verbatim}
\usepackage{siunitx}
\usepackage[italic]{hepnames}
\usepackage{color}
\usepackage{cancel}

\begin{document}

\preprint{FERMILAB-PUB-22-527-ND-T, UMN-TH-4128/22, FTPI-MINN-22/19}

\title{First Constraints on Heavy QCD Axions with a Liquid Argon Time Projection Chamber using the ArgoNeuT Experiment}

\author{R.~Acciarri}
\affiliation{Fermi National Accelerator Laboratory, Batavia, Illinois 60510, USA}

\author{C.~Adams}
\affiliation{Argonne National Laboratory, Lemont, Illinois 60439, USA}

\author{B.~Baller}
\affiliation{Fermi National Accelerator Laboratory, Batavia, Illinois 60510, USA}

\author{V.~Basque}
\affiliation{Fermi National Accelerator Laboratory, Batavia, Illinois 60510, USA}

\author{F.~Cavanna}
\affiliation{Fermi National Accelerator Laboratory, Batavia, Illinois 60510, USA}

\author{R.~T.~Co}
\affiliation{School of Physics and Astronomy, University of Minnesota, Minneapolis, MN 55455, USA}
\affiliation{William I. Fine Theoretical Physics Institute, University of Minnesota, Minneapolis, MN 55455, USA}

\author{R.~S.~Fitzpatrick}
\affiliation{University of Michigan, Ann Arbor, Michigan 48109, USA}

\author{B.~Fleming}
\affiliation{Yale University, New Haven, Connecticut 06520, USA}

\author{P.~Green}
\email{patrick.green@physics.ox.ac.uk}
\affiliation{University of Manchester, Manchester M13 9PL, United Kingdom}
\affiliation{University of Oxford, Oxford OX1 3RH, United Kingdom}

\author{R.~Harnik}
\affiliation{Fermi National Accelerator Laboratory, Batavia, Illinois 60510, USA}

\author{K.~J.~Kelly}
\affiliation{CERN, Esplande des Particules, 1211 Geneva 23, Switzerland}

\author{S.~Kumar}
\affiliation{University of California, Berkeley, California 94720, USA}
\affiliation{Lawrence Berkeley National Laboratory, Berkeley, California 94720, USA}

\author{K.~Lang}
\affiliation{University of Texas at Austin, Austin, Texas 78712, USA}

\author{I.~Lepetic}
\affiliation{Rutgers University, Piscataway, New Jersey 08854, USA}

\author{Z.~Liu}
\email{zliuphys@umn.edu}
\affiliation{School of Physics and Astronomy, University of Minnesota, Minneapolis, MN 55455, USA}

\author{X.~Luo}
\affiliation{University of California, Santa Barbara, California, 93106, USA}

\author{K.~F.~Lyu}
\affiliation{School of Physics and Astronomy, University of Minnesota, Minneapolis, MN 55455, USA}

\author{O.~Palamara}
\affiliation{Fermi National Accelerator Laboratory, Batavia, Illinois 60510, USA}

\author{G.~Scanavini}
\affiliation{Yale University, New Haven, Connecticut 06520, USA}

\author{M.~Soderberg}
\affiliation{Syracuse University, Syracuse, New York 13244, USA}

\author{J.~Spitz}
\affiliation{University of Michigan, Ann Arbor, Michigan 48109, USA}

\author{A.~M.~Szelc}
\affiliation{University of Edinburgh, Edinburgh EH9 3FD, United Kingdom}

\author{W.~Wu}
\affiliation{Fermi National Accelerator Laboratory, Batavia, Illinois 60510, USA}

\author{T.~Yang}
\affiliation{Fermi National Accelerator Laboratory, Batavia, Illinois 60510, USA}

\collaboration{The ArgoNeuT Collaboration}
\noaffiliation

\begin{abstract}
We present the results of a search for heavy QCD axions performed by the ArgoNeuT experiment at Fermilab. We search for heavy axions produced in the NuMI neutrino beam target and absorber decaying into dimuon pairs, which can be identified using the unique capabilities of ArgoNeuT and the MINOS near detector. This decay channel is motivated by a broad class of heavy QCD axion models that address the strong CP and axion quality problems with axion masses above the dimuon threshold. We obtain new constraints at a 95\% confidence level for heavy axions in the previously unexplored mass range between 0.2-0.9~GeV, for axion decay constants around tens of TeV.
\end{abstract}

\maketitle

{\bf \textit{Introduction}}.---The QCD axion was proposed~\cite{Peccei:1977hh,Peccei:1977ur} to address the strong CP problem~\cite{tHooft:1976rip, Baker:2006ts, Afach:2015sja, Graner:2016ses, Abel:2020gbr}. However, in the simplest implementations, this mechanism suffers from the axion quality problem~\cite{Giddings:1988cx, Coleman:1988tj, Gilbert:1989nq}. Heavy QCD axions, defined as those with a coupling to gluons but with a much larger mass than the QCD axion, are motivated by their potential role resolving the axion quality problem while preserving the solution to the strong CP problem~\cite{Rubakov:1997vp,Berezhiani:2000gh,Hook:2014cda,Fukuda:2015ana,Gherghetta:2016fhp,Agrawal:2017ksf,Hook:2019qoh,Gherghetta:2020keg}. Furthermore, they can explain various phenomena in astrophysics~\cite{Nomura:2008ru} and cosmology~\cite{Takahashi:2021tff,Co:2022aav}. In these models, the larger axion mass, $m_a$, and smaller axion decay constant, $f_a$, also open up various decay channels involving Standard Model (SM) particles and enhance the axion interaction strengths. This enables searches for these heavy QCD axions in beam-dump and collider experiments.

In this \emph{Letter}, we perform a search for heavy QCD axions with $200~\text{MeV}\lesssim m_a \lesssim 1~\text{GeV}$ using the ArgoNeuT experiment~\cite{Anderson:2012vc} -- a 0.24 ton Liquid Argon Time Projection Chamber (LArTPC) neutrino detector that collected five months of data in 2009-2010 in the Neutrinos at the Main Injector (NuMI) beamline~\cite{Adamson:2015dkw} at Fermilab. The data used corresponds to \num{1.25e20} protons-on-target (POT) acquired while the NuMI beam was in anti-neutrino mode~\cite{Anderson:2012vc}. The axions can be produced via couplings with SM mesons and protons when the 120 GeV proton beam strikes the graphite target or the hadron absorber located 1033\,m and 318\,m upstream of ArgoNeuT, respectively. The produced axions can then propagate to ArgoNeuT where the decay signature $a \to \mu^+ \mu^-$ is searched for. The muon pair is reconstructed in ArgoNeuT as minimally ionizing tracks that can then be matched to a pair of tracks with opposite charges in the magnetised MINOS near detector (MINOS-ND)~\cite{Michael:2008bc} located immediately downstream of ArgoNeuT. 

{\bf \textit{Heavy QCD axions}}.---
Heavy QCD axions must couple to gluons to solve the strong CP problem.
Furthermore, consistent with Grand Unified Theories, these axions can also couple to the other gauge bosons of the SM. These considerations motivate the following couplings~\cite{Bauer:2017ris},
\begin{align}\label{eq:lag_gauge}
\mathcal{L}_{\rm gauge} = \frac{c_3\alpha_3}{8\pi f_a}a G\tilde{G} + \frac{c_2\alpha_2}{8\pi f_a}a W\tilde{W} + \frac{c_1\alpha_1}{8\pi f_a}a B\tilde{B}.
\end{align}
Here $G$ is SM gluon field strength and $\tilde{G}$ is its dual. Couplings to $SU(2)$ and $U(1)$ gauge fields, $W\tilde{W}$ and $B\tilde{B}$, are defined analogously. The coefficients $\alpha_i = g_i^2/(4\pi)$ are given in terms of the three gauge couplings $g_i$ defined at the scale $m_a$. We will set $c_3=c_2=c_1=1$ hereafter. We note, however, that the results presented in this \emph{Letter} primarily depend on the $aG\tilde{G}$ coupling and would still apply if $c_1, c_2 \ll c_3$.

Along with couplings to gauge bosons, axions can also couple to SM fermions, as in the DFSZ models~\cite{Dine:1981rt,Zhitnitsky:1980tq}. While coupling to both quarks and leptons can appear, we consider axion coupling to only SM leptons in a flavor diagonal way in order to focus on a parameter space that is complementary to the multitude of flavor searches (see, e.g.,~\cite{Goudzovski:2022vbt} for a recent summary and references) and is theoretically well motivated. Therefore, we consider~\cite{Bauer:2017ris}
\begin{align}\label{eq:lag_lepton}
\mathcal{L}_{\rm lepton} = \sum_{\ell=e,\mu,\tau}\frac{\partial_\mu a}{2f_a}\left(c_{V\ell}\bar{\ell}\gamma^{\mu}\ell + c_{A\ell}\bar{\ell}\gamma^{\mu}\gamma_5\ell\right).
\end{align}
Here $c_{V\ell},c_{A\ell}$ control the flavor universal vector and axial coupling of the axion to SM charged leptons.

\begin{figure}[tb]
    \centering
    \includegraphics[width = 0.5\textwidth]{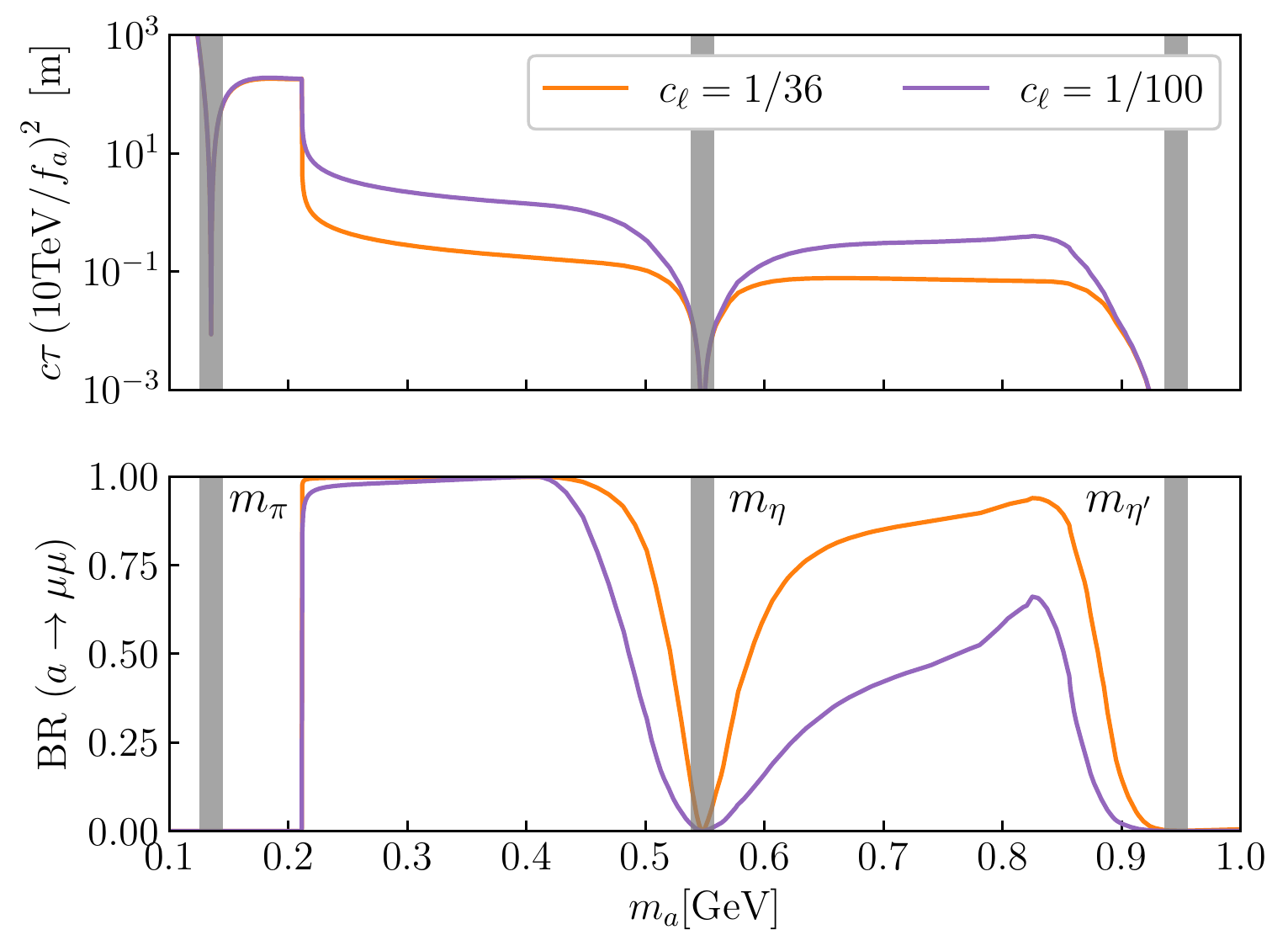}
    \caption{Lifetime (top) and dimuon branching ratio (bottom) of the axion for the two benchmark scenarios as a function of the axion mass. 
    }
    \label{fig:lifetime}
\end{figure}

Given the axion couplings in Eqs.~\eqref{eq:lag_gauge} and \eqref{eq:lag_lepton}, the relevant decay modes of the axions for masses $0.2~\text{GeV}< m_a <1~\text{GeV}$ are into photons, muons and some exclusive hadronic modes. Considering these decay modes, we show the behaviour of the decay length of the axion in its rest frame for $f_a=10$~TeV in Fig.~\ref{fig:lifetime} (top). Discussion of the individual contributions of the various decay channels involved can be found in the Supplemental Material~\cite{supp}. Hereafter, we focus on two theory benchmarks with $c_\ell\approx c_{A\ell} = 1/36$ and $c_\ell\approx c_{A\ell}=1/100$ to illustrate the constraining power in the $f_a$-$m_a$ plane. Choosing $c_\ell$ smaller than $c_i=1$ is motivated because, from a theoretical perspective, a suppressed leptonic coupling can naturally emerge in models where axions directly couple to some new heavy leptons, which in turn mix with SM leptons giving $c_{\ell} \propto \theta_{\rm mix}^2 \ll 1$~\cite{Kim:2004rp, Kaplan:2015fuy, Buen-Abad:2021fwq}. Our choice of $c_{\ell}=1/36$ and $c_{\ell}=1/100$ then corresponds to small mixing angle benchmarks $\theta_{\rm mix} \approx 1/6$ and $\theta_{\rm mix} \approx 1/10$, respectively, that can arise from loop-induced or vector-like fermion-induced models~\cite{Co:2022bqq}. From the perspective of axion searches in ArgoNeuT, a smaller $c_\ell < c_i$ also makes the axion sufficiently long-lived to reach the detector while not suppressing its production via the gluon coupling.

Since our search is based on muon final states, we also show in Fig.~\ref{fig:lifetime} (bottom) the branching ratio of the axion into two muons for the two benchmarks. The dimuon mode is a dominant decay channel for most of the mass range, enabling ArgoNeuT to be uniquely sensitive to these scenarios. In regions where dimuon decays are subdominant, future searches in other channels, e.g. $\gamma\gamma$ and multi-hadron states, could provide complimentary coverage.

{\bf \textit{Generation and simulation}}.---
Due to the presence of the gluon coupling, axions mix with SM pseudoscalar mesons $\pi,\eta,\eta'$~\footnote{Other modes of axion production, such as from meson decays, are subdominant unless for example additional flavor-violating couplings are introduced. Such models are not considered in this analysis. A recent work~\cite{Bertuzzo:2022fcm} focused on such a model, but without a gluon coupling.} with the corresponding mixing angles $\theta_{a M}$, $M=\pi,\eta,\eta'$, given by~\cite{Bauer:2017ris,Ertas:2020xcc,Kelly:2020dda}
\begin{equation}\label{eq:mixing}
\begin{aligned}
\theta_{a M} = \dfrac{f_\pi}{f_a}\frac{C_{aM}}{m_a^2-m_M^2}.
\end{aligned}
\end{equation}
Here $f_\pi = 93$~MeV is the pion decay constant and $C_{a\pi} = m_a^2/6$, $C_{a\eta} = (m_a^2- 4 m_\pi^2/9)/\sqrt{6}$ and $C_{a\eta'} = (m_a^2 - 16 m_\pi^2/9)/(2 \sqrt{3})$. Our subsequent analysis assumes that $\theta_{a M}\ll 1$, and therefore we mask the parameter space for which $m_a\approx m_M$, as shown by vertical gray bands in Fig.~\ref{fig:lifetime}. To compute the spectrum of axions produced in the NuMI beam we first simulate the spectrum of SM mesons using \texttt{Pythia8}~\cite{Bierlich:2022pfr,Sjostrand:2014zea}. We find that on average 2.89\,$\pi^0$, 0.33\,$\eta$ and 0.03\,$\eta^\prime$ are produced per proton collision. Subsequently, applying the mixing angles in Eq.~\eqref{eq:mixing}, we can compute the total number of axions produced considering both the 87\% of protons interacting in the NuMI target and the 10\% that reach the downstream hadron absorber with energies $\sim$120\,GeV~\cite{Adamson:2015dkw}. Using the geometrical acceptance of ArgoNeuT and the axion branching ratio into muons (Fig.~\ref{fig:lifetime}), we then compute the axion decays to a dimuon final state that would be seen in the detector. The LArSoft software framework~\cite{Snider:2017wjd} is used to simulate these muons in ArgoNeuT. LArSoft propagates the particles using GEANT4 \cite{Agostinelli:2002hh} and then performs detector response modelling and reconstruction \cite{Anderson:2012vc,Acciarri:2018ahy}. The standard MINOS simulation and reconstruction framework is then used to model the particles that exit ArgoNeuT and reach the MINOS-ND~\cite{Anderson:2012vc, Michael:2008bc}.

\begin{figure}[tb]
    \centering
    \includegraphics[width = 0.5\textwidth]{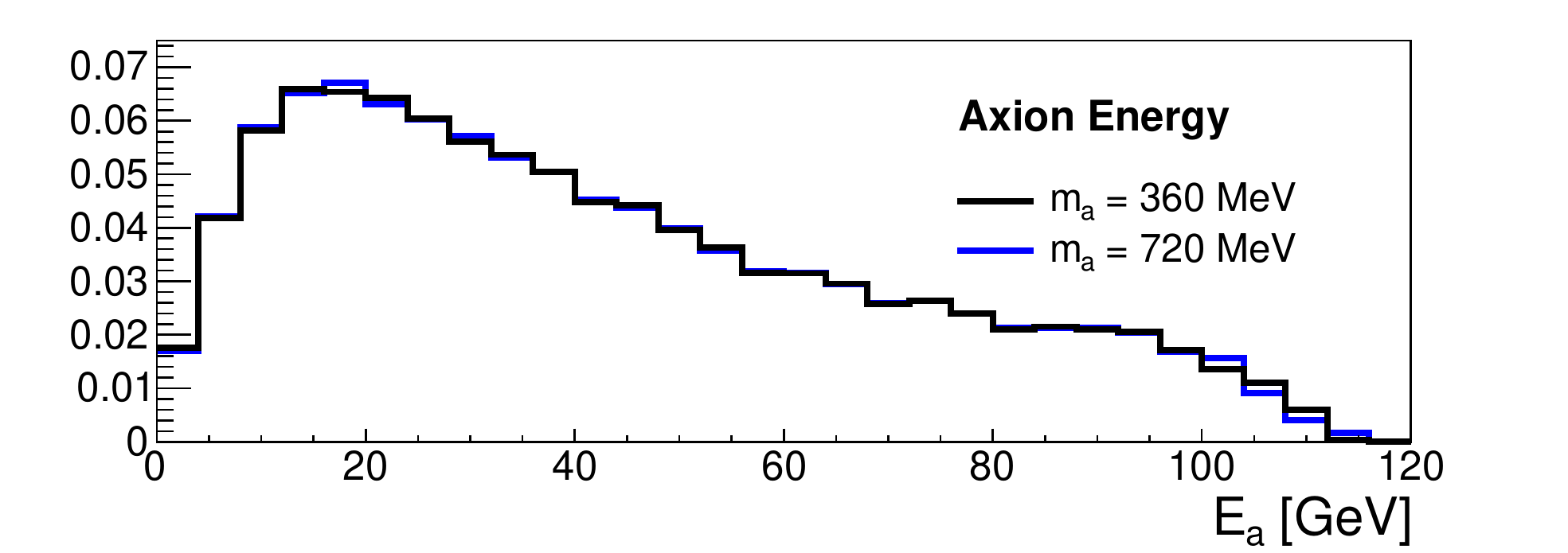}
    \includegraphics[width = 0.5\textwidth]{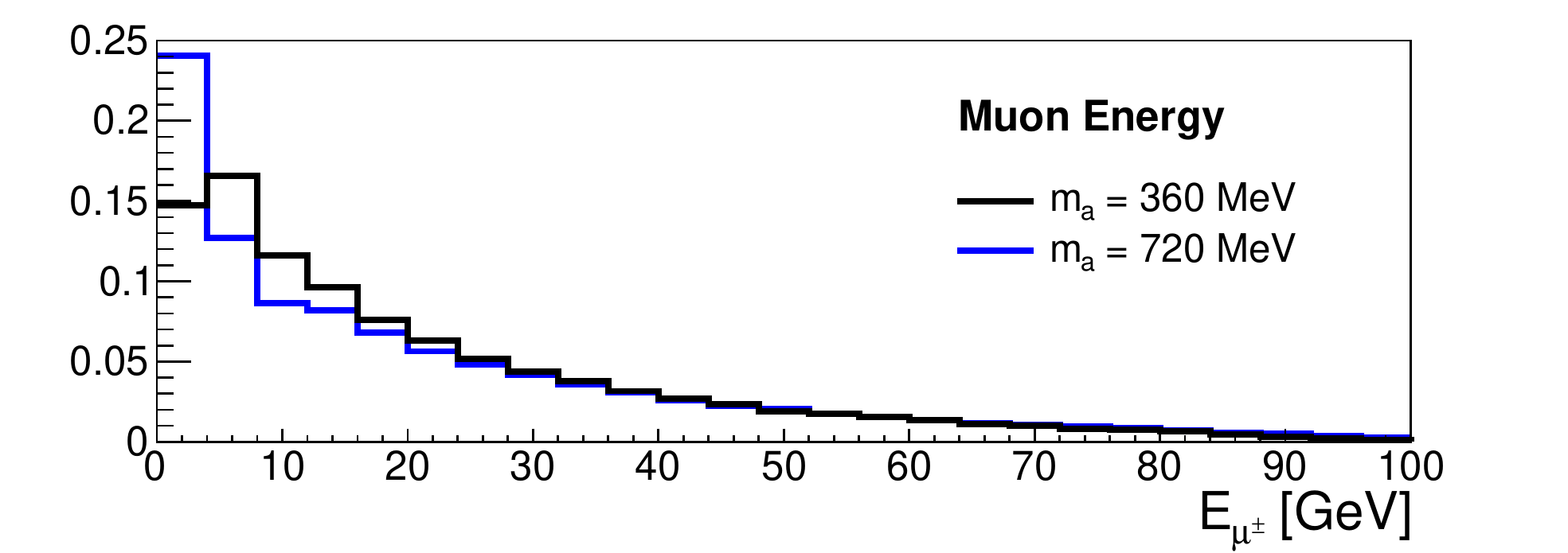}
    \includegraphics[width = 0.5\textwidth]{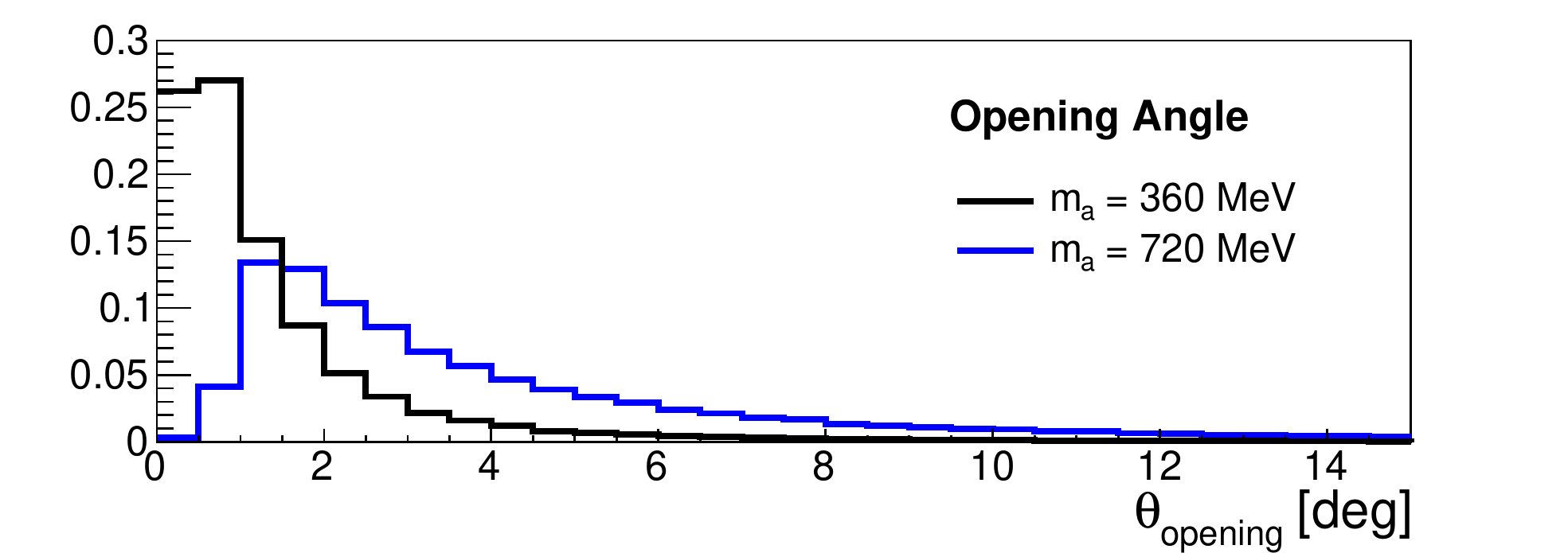}
    \caption{Parent axion energy (top), energy of the resulting individual muons (middle) and opening angle between them (bottom) for axions with $m_a = 360$\,MeV (black) and $m_a = 720$\,MeV (blue).}
    \label{fig:kinematics}
\end{figure}

{\bf \textit{Signature and selection}}.--- 
In ArgoNeuT the axion decay $a \rightarrow \APmuon \Pmuon$ can be detected as a pair of minimally ionizing particles (MIPs). The parent axion energies and the kinematics of the resulting muons for two different axion masses are shown in Fig.~\ref{fig:kinematics}. The muons are highly energetic, with $\langle E_{\mu^\pm} \rangle \approx 20 $\,GeV, resulting in them typically exiting ArgoNeuT and propagating to the downstream MINOS-ND. They are also highly forward-going with an average angle with respect to the beam direction of $\langle \theta_{\rm beam} \rangle \approx 0.75 \degree$ to $\langle \theta_{\rm beam} \rangle \approx 2.5 \degree$ and an average opening angle between them of $\langle \theta_{\rm opening} \rangle \approx 1.5 \degree$ to $\langle \theta_{\rm opening} \rangle \approx 5.0 \degree$, in each case depending on $m_a$. As a result of ArgoNeuT's angular reconstruction resolution of between $1\degree$ and $3\degree$ (depending on track orientation)~\cite{spitzthesis}, the muons may overlap and appear as one track in ArgoNeuT. Once reaching the MINOS-ND, the muons then separate due to their opposite charges in the MINOS-ND magnetic field of about 1\,T~\cite{Michael:2008bc}.

The axion decay signature is similar to the signature in ArgoNeuT's previous heavy neutral lepton search \cite{ArgoNeuT:2021clc} and an analogous selection strategy can be followed. However, it is a resonant two-body decay as opposed to the non-resonant three-body decay previously considered. Therefore, given the different nature of this decay and the different production mechanism and resulting kinematics, the selection has been modified and re-optimised. We consider two different scenarios, depending on how forward-going the muon pair are. In the first, the pair of muons are sufficiently separated to be reconstructed as two distinct minimally ionizing tracks. In the second, the separation between them is too small and they are instead reconstructed as a single track with twice the minimally ionizing particle $dE/dx$. These will be referred to subsequently as {\it two-track} and {\it double-MIP} type events, respectively. Due to the highly forward-going nature of the muon pairs resulting from axion decays, the {\it double-MIP} signature dominates. In both cases, the tracks in ArgoNeuT can be matched with two tracks in the downstream MINOS-ND that are reconstructed with opposite charges. In the second scenario, axion decays occurring in the upstream cavern are also considered. The resulting muons then pass through the ArgoNeuT detector and can be matched to the MINOS-ND as before. The ArgoNeuT physics run coincided with the construction of the upstream MINERvA detector~\cite{Aliaga:2013uqz}. Therefore, only the 63\,cm region between MINERvA and ArgoNeuT is taken into account. 

A series of pre-selection cuts are first applied. Events with incomplete reconstruction are removed by requiring that at least 80\% of energy depositions are associated with reconstructed objects. Next, events are identified that are compatible with axion decays: requiring a maximum of two tracks in ArgoNeuT, where short tracks ($L \leq 5$\,cm) are ignored to ensure events are not removed due to the presence of $\delta$-rays, and at least two tracks in the MINOS-ND. Reconstructed tracks are also classified based on whether they originate from within or outside of the ArgoNeuT fiducial volume: $1 \leq x \leq 46$\,cm (drift), $-19 \leq y \leq 19$\,cm (vertical) and $z \geq 3$\,cm (beam direction).

Two selection paths are then followed in ArgoNeuT depending on the event topology. In the two-track case events are selected that have two minimally-ionising tracks originating from a common vertex within the ArgoNeuT fiducial volume. Tracks are considered to be minimally-ionising if they have an average $dE/dx < 3.1$\,MeV/cm and are considered to have originated from a common vertex if they have a separation of less than 4\,cm between the track starts. The pair of tracks are also required to have an opening angle between them of $\theta_{opening} \leq 15\degree$ and exit towards the MINOS-ND. In the double-MIP case events are selected that have a single track, originating from either within ArgoNeuT or from upstream of the detector, that has an average $dE/dx$ consistent with two overlapping minimally ionising particles. The region near the track vertex prior to the potential separation of the two muons is assessed: considering the average $dE/dx$ over the first 5\,cm, where any high energy depositions ($dE/dx > 10$\,MeV/cm) from $\delta$-rays are removed. The track is required to have an average $dE/dx > 3.1$\,MeV/cm. Additionally, the track is required to be forward-going with an angle with respect to the beam direction $\theta_{beam} \leq 10\degree$ and exit towards the MINOS-ND. 

Next, matching to the MINOS-ND is performed. In the two-track scenario each of the two tracks are individually matched to tracks reconstructed in the MINOS-ND, with matching tolerances of $r_{\rm diff} \leq 12.0$\,cm (radial) and $\theta_{\rm diff} \leq 0.17$\,rad (angular) accounting for the expected maximum deflection of the muons between the two detectors~\cite{Acciarri:2018ahy}. The matched tracks are also both required to start within the calorimeter region of the MINOS-ND, which is directly downstream of the ArgoNeuT detector, and start within $20$\,cm of the first instrumented plane. In the double-MIP scenario, since a single track in ArgoNeuT now being matched to two tracks in the MINOS-ND, the matching tolerances are loosened to double the two-track case. Several further selection cuts are applied to the matched tracks in the MINOS-ND. They are required to both be muon-like based both on the track length and $dE/dx$: $L_{MINOS-ND} \geq 1$\,m and $6 \leq dE/dx_{MINOS-ND} \leq 12$\,MeV/cm. They are also required to have timestamps consistent with having originated from a single decay in ArgoNeuT, $|\Delta t_{0}| < 20$\,ns, and be reconstructed with opposite charges.

\begin{figure}
  \centering
  \includegraphics[width=.5\textwidth,keepaspectratio]{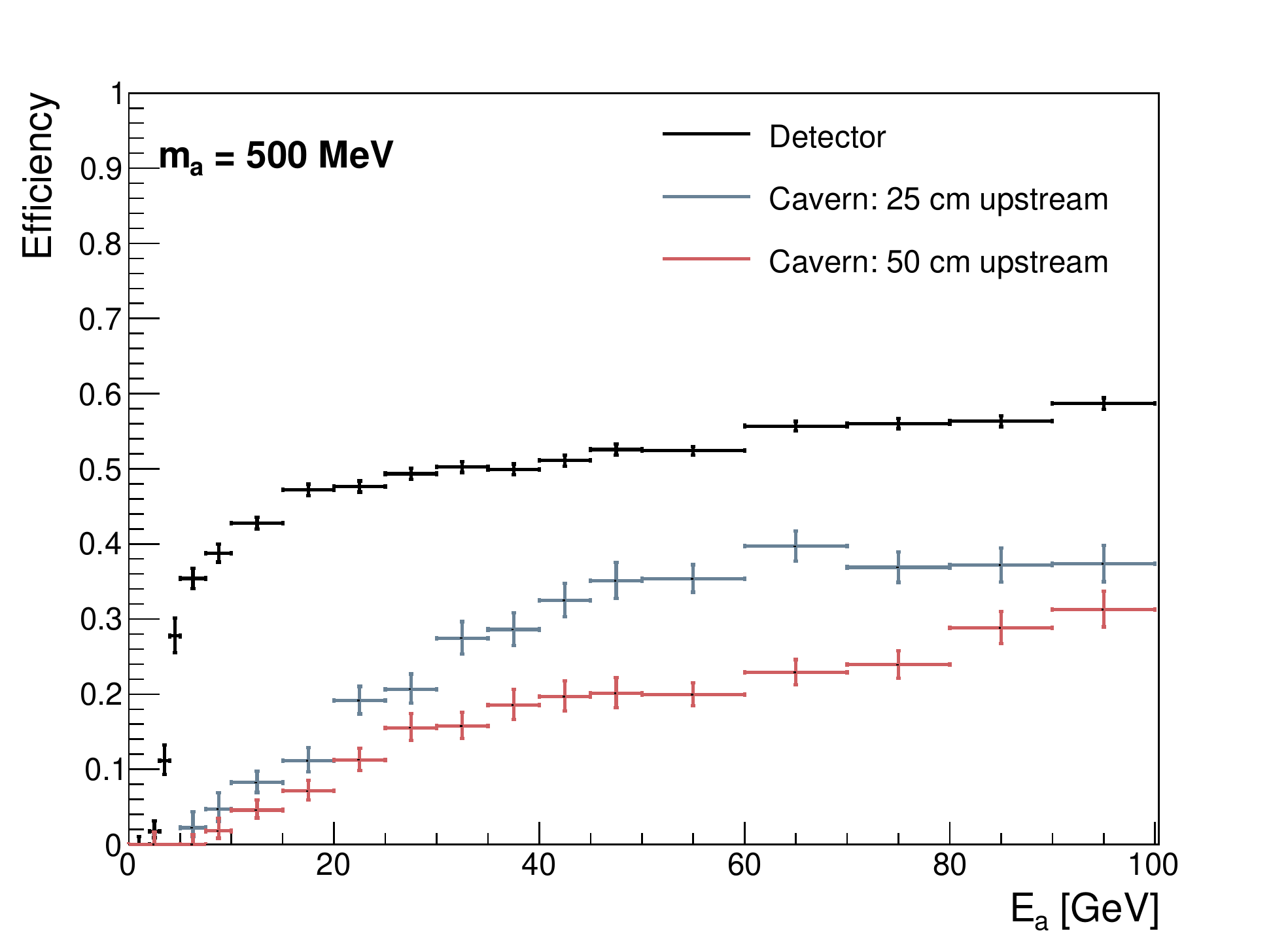}
  \caption{Selection efficiency for $m_a=500$\,MeV axion decays as a function of axion energy, $E_a$. Decays occurring inside ArgoNeuT (black) and at 25\,cm (blue) and 50\,cm (red) upstream of the detector are shown.}
  \label{fig:efficiency}
\end{figure}

The selection efficiency is shown in Fig.~\ref{fig:efficiency} for simulated decays of axions with mass $m_a=500$\,MeV as a function of the axion energy, $E_a$. Decays occurring inside ArgoNeuT (black) and at 25\,cm (blue) and 50\,cm (red) upstream of the detector are shown. For decays inside the detector the efficiency is around 50\% above $E_a \sim 15$\,GeV. However, at lower $E_a$ one or both of the muons may have insufficient energy to propagate to the MINOS-ND before stopping. This causes the matching to fail resulting in a sharp decline in the selection efficiency. For decays upstream of the detector, where only the double-MIP topology is considered, the selection efficiency is lower. As the distance the muons have to propagate before reaching the detector increases, the less likely they are to remain overlapping resulting in the events being rejected. At lower $E_a$ the muons are less boosted, further increasing the likelihood they separate.

For selected axion candidate events the invariant mass of the parent axion can be reconstructed. This is achieved using the trajectories of the tracks reconstructed in ArgoNeuT (or matching between ArgoNeuT and the MINOS-ND in the double-MIP scenario), combined with the momentum reconstruction in the MINOS-ND. The axion invariant mass can be reconstructed with a resolution of $\sim$100\,MeV when both muons are contained within the MINOS-ND, and $\sim$200\,MeV if exiting. A constraint on the invariant mass is not applied in the selection since the search is performed across a significant axion mass range. However, if a signal were to be observed in the data the invariant mass could be a powerful tool to further characterise it.  

{\bf \textit{Backgrounds and systematic uncertainties}}.---

The primary backgrounds to this search originate from beam neutrino interactions within the detector or the surrounding materials. These are modelled using the GENIE neutrino generator~\cite{Andreopoulos:2009rq} along with a data-driven model of the rate and kinematics of beam-induced through-going muons~\cite{spitzthesis, ArgoNeuT:2011bms, Acciarri:2014isz}. Modelling of the NuMI beam flux developed by the MINERvA collaboration is used~\cite{Aliaga:2016oaz}. The most significant backgrounds arise from charged-current muon neutrino interactions: either single charged pion production, resulting in two approximately minimally ionising tracks that could be mistaken for the two-track signature; or quasi-elastic scattering producing a single muon with low energy protons near the interaction vertex causing it to mimic the double-MIP signature. In each case, these could then be incorrectly matched with other nearby muons in the MINOS-ND. However, the majority of these types of interactions are removed during the selection due to either the event topology in ArgoNeuT or the MINOS-ND or as a result of the precise timing resolution of the MINOS-ND~\cite{Michael:2008bc}. The total background expectation for the ArgoNeuT data-set is $0.1 \pm 0.1$ events. 

The systematic uncertainties are dominated by theoretical uncertainties in the axion flux prediction. One way to estimate these is from the strong coupling, $\alpha_S$, uncertainty of around $20\%$~\cite{Deur:2016tte}. Another way is from the uncertainty on the measured pion flux of the NuMI beam that is at the 5-10\% level~\cite{MIPP:2014shj}. Furthermore, the production is modeled by chiral perturbation theory where the matched effective coefficients also have uncertainties. The current best fit for the pion coupling has an uncertainty of 10-20\%~\cite{Borasoy:1998pe}. To conservatively illustrate the impact of these uncertainties, the theory prediction of the axion flux is varied by 30\% in the final sensitivity evaluation. Future refinements in the axion flux modeling would improve the accuracy of the derived limit. The impact of experimental systematic uncertainties is also evaluated. These arise from uncertainties in the muon reconstruction and resulting selection efficiency in ArgoNeuT (3.3\%)~\cite{Acciarri:2018ahy, spitzthesis, Anderson:2012vc, Anderson:2012mra} and the MINOS-ND (0.4\%)~\cite{Adamson:2009ju}, the POT evaluation (1\%)~\cite{Acciarri:2014isz}, and the determination of the electron drift velocity and hence the total volume of instrumented argon (2.2\%)~\cite{spitzthesis}. The experimental systematic uncertainties have a combined impact of 4.1\%. 

\begin{figure}[tb]
\centering
\includegraphics[width = 0.48\textwidth]{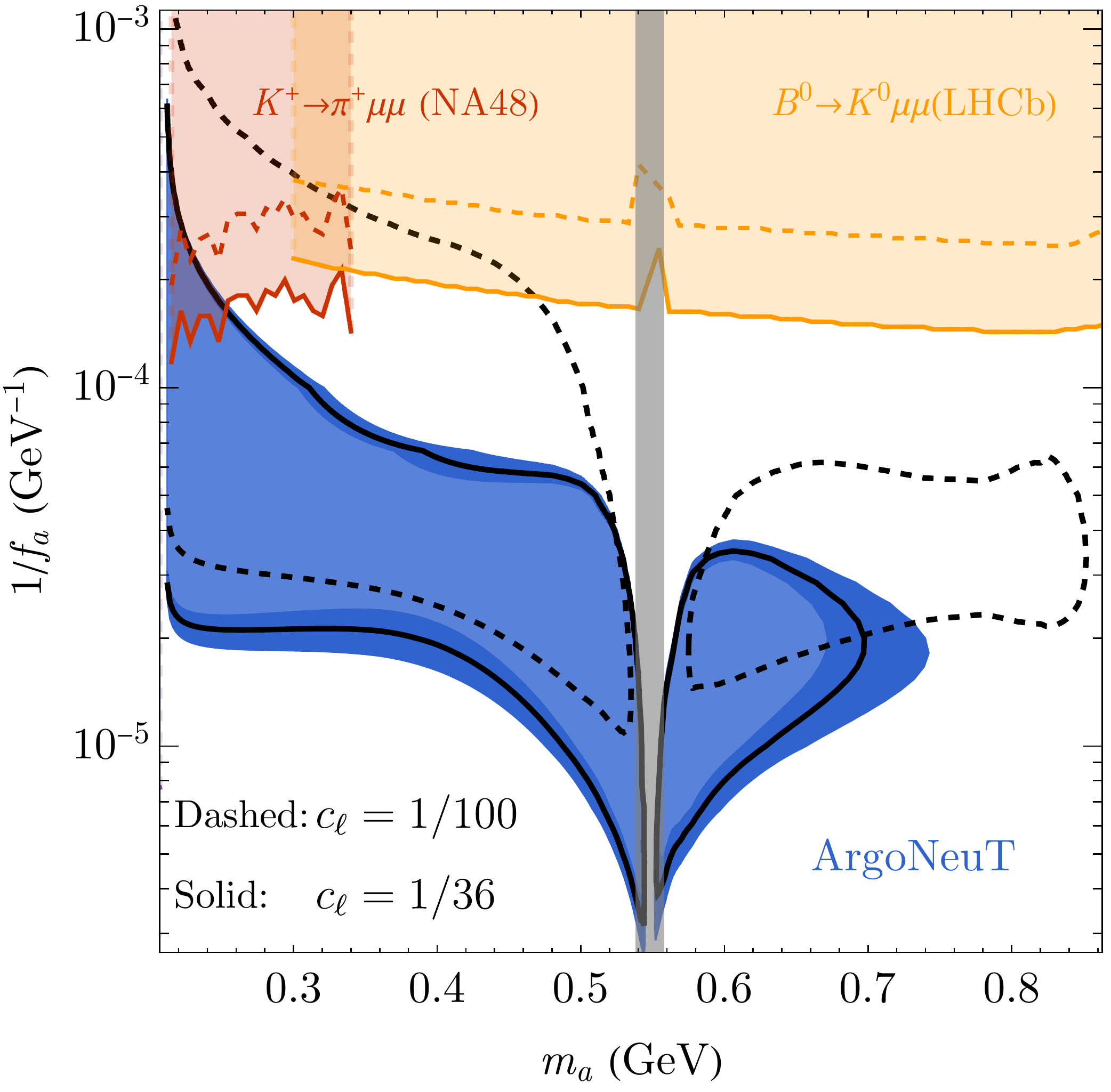}
\caption{Constraints on the axion model parameter space at 95\% CL from the exposure of $1.25 \times 10^{20}$ POT at ArgoNeuT (blue shaded region and black contours). The derived limits for $c_\ell=1/36$ and $c_\ell=1/100$ are shown by the solid and dashed contours. The uncertainty on the expected constraint, predominantly arising from the theoretical uncertainties, is shown by the dark blue band. The red and orange contours show the strongest existing constraints evaluated for the two benchmark scenarios. The gray-shaded band indicates a region with increased theoretical uncertainty around the $\eta$ mass.}
\label{fig:results}
\end{figure}

{\bf \textit{Results}}.---
Zero events pass the selection in ArgoNeuT's full \num{1.25e20}~POT anti-neutrino-mode data-set, consistent with the background prediction of $0.1 \pm 0.1$ events. Our exclusion of parameter space at a 95\% confidence level is shown in Fig.~\ref{fig:results} for both the $c_\ell=1/36$ and $c_\ell=1/100$ scenarios, evaluated using a Bayesian approach with a uniform prior~\cite{PDG:2022}. The impact of the uncertainties on the expected constraint, dominated by the theoretical uncertainty, is also shown for the $c_\ell=1/36$ case~\footnote{The impact of the uncertainties on the $c_\ell=1/100$ case is similar, but is not shown to aid clarity.}. In the presence of an axion there are new decay modes for SM mesons, such as $B^+\rightarrow K^+ a$, where $a$ can subsequently decay into $\mu\mu$. Searches for such rare decay modes place important constraints on our parameter space. We find the following searches give significant constraints: $K^+\rightarrow \pi^+ \nu\bar{\nu}$ by the NA62 collaboration~\cite{NA62:2021zjw}; $K^{\pm}\rightarrow \pi \mu\mu$ by the NA48/2 collaboration~\cite{NA482:2016sfh}; $B^0\rightarrow K^{*0}\mu\mu$ by the LHCb collaboration~\cite{LHCb:2015nkv}; and $B^+\rightarrow K^{+}\eta\pi\pi$ by the BaBar collaboration~\cite{BaBar:2008rth}. For all these cases, we recast the presented bounds as appropriate for the axion lifetime in our scenario. The strongest resulting constraints are shown in Fig.~\ref{fig:results} for each benchmark model. 

This measurement leads to new constraints on previously unexplored parameter space for heavy QCD axions with masses above the dimuon threshold and below 1~GeV (where hadronic decays would dominate). The coverage of the axion decay constant for the benchmark model with $c_\ell=1/36$ is around $f_a\sim 50$~TeV for masses up to 0.65~GeV. The coverage of the axion decay constant for the benchmark model with $c_\ell=1/100$ is around $f_a\sim 20$~TeV for masses up to 0.84~GeV. For these benchmark couplings ArgoNeuT provides significant improvement on existing constraints. ArgoNeuT also has constraining power for couplings ranging between approximately $c_\ell = \mathcal{O}(10^{-1}\mathchar`-10^{-3})$. For much larger $c_\ell$ the axions would predominantly decay before reaching the detector reducing the constraining power, and for much smaller $c_\ell$ the axions would no longer dominantly decay to muons but to mesons and photons that are not searched for in this analysis. To explore these regions of phase space would require future searches with detectors at shorter baselines than ArgoNeuT, or probing different decay modes.

{\bf \textit{Conclusions}}.--- We have presented the first search for heavy QCD axions in a LArTPC using the ArgoNeuT experiment. This type of axion is particularly motivated by the strong CP puzzle and the axion quality problem. We search for such axions produced in the NuMI neutrino beam and then decaying with a dimuon signature within, or close to, the ArgoNeuT detector. In ArgoNeuT's \num{1.25e20} POT anti-neutrino mode data-set zero events pass the selection, consistent with the background prediction. This measurement leads to a significant new exclusion region for heavy axions in the mass range between 0.2-0.9~GeV for an axion decay constant around tens of TeV, over a broad range of axion-lepton couplings around $c_\ell = \mathcal{O}(10^{-1}\mathchar`-10^{-3})$. The search can be extended to various new heavy QCD axion models and paves the way for heavy QCD axion searches at future neutrino facilities. The techniques developed could also be used to constrain other dark sector particle models with long-lived resonance decays into dimuons.

{\bf \textit{Acknowledgements}}.---
This manuscript has been authored by Fermi Research Alliance, LLC under Contract No. DE-AC02-07CH11359 with the U.S. Department of Energy, Office of Science, Office of High Energy Physics. We gratefully acknowledge the cooperation of the MINOS Collaboration in providing their data for use in this analysis. We wish to acknowledge the support of Fermilab, the Department of Energy, and the National Science Foundation in ArgoNeuT’s construction, operation, and data analysis. This project has received funding from the Science and Technology Facilities Council (STFC), part of the United Kingdom Research and Innovation; and from the Royal Society UK grants: RGF\textbackslash EA\textbackslash 180209 and URF\textbackslash R\textbackslash 201022. R.C. was supported in part by U.S. Department of Energy (DOE) grant DE-SC0011842. S.K. was supported in part by the National Science Foundation (NSF) grant PHY-1915314 and the DOE contract DE-AC02-05CH11231. Z.L. and K.F.L. were supported in part by the DOE grant DE-SC0022345. For the purpose of Open Access, the author has applied a CC BY public copyright licence to any Author Accepted Manuscript version arising from this submission.

\bibliography{main.bib}

\end{document}


\title{Supplemental material}
\maketitle

\section{Summary of decay modes}
\paragraph{Decay into muons.}
The axion decay width into muons is given by,
\begin{align}
\Gamma_{a\rightarrow \mu\mu} = \frac{c^2_{\ell} m_a m_\mu^2}{8\pi f_a^2} \sqrt{1-\frac{4m_\mu^2}{m_a^2}}.    
\end{align}
The parameter $c_\ell$ is determined by $c_1, c_2, c_3, c_{V\ell}, c_{A\ell}$, but it is a good approximation to use $c_\ell\approx c_{A\ell}$, defined in the main text. The axion decay width into electrons compared to muons is suppressed by $m_e^2/m_\mu^2$, and hence will not be important for the mass range we consider in this work.
\paragraph{Decay into photons.}
The axion decay width into two photons is given by
\begin{align}
\Gamma_{a\rightarrow \gamma\gamma} = \frac{\alpha_{\rm em}^2 |c_\gamma|^2 m_a^3}{256\pi^3 f_a^2}.
\end{align}
Here $c_\gamma$ is the effective photon coupling given in terms of $c_1, c_2, c_3, c_{V\ell}, c_{A\ell}$. One of the contributions to $c_\gamma$ is from $a W\tilde{W}$ and $a B\tilde{B}$ couplings after electroweak symmetry breaking. It also gets a contribution from various hadronic modes since the axion can mix with Standard Model (SM) pseudoscalar mesons $\pi,\eta,\eta'$ via the axion-gluon $a G\tilde{G}$ coupling. Finally, it also receives a contribution at one-loop from the tree level lepton couplings that is present in the class of axion models considered in this work. These contributions can be summarized as,
\begin{equation}
\begin{aligned}
c_\gamma & = c_2+\frac{5}{3} c_1 + c_3\left(-1.92+\frac{1}{3}\frac{m_a^2}{m_a^2-m_\pi^2}+\frac{8}{9}\frac{m_a^2-\frac{4}{9}m_\pi^2}{m_a^2-m_\eta^2}+\frac{7}{9}\frac{m_a^2-\frac{16}{9}m_\pi^2}{m_a^2-m_{\eta'}^2}\right) + 2 \sum_{\ell=e,\mu,\tau} c_{A\ell} B_1(4m_l^2/m_a^2).
\label{eq.cgalow}
\end{aligned}
\end{equation}
Here $B_1(x) = 1- x  g(x)^2$ and 
\begin{align}
g(x)=
\begin{cases}
\sin^{-1}\left(\frac{1}{\sqrt{x}}\right)    & \quad \text{if }  x\geq 1\\
    \frac{\pi}{2} + \frac{i}{2} \log\left(\frac{1+\sqrt{1-\tau}}{1-\sqrt{1-\tau}}\right) & \quad \text{if } x<1.
\end{cases}
\end{align}

\begin{figure}[htbp]
    \centering
    \includegraphics[width=0.48\textwidth]{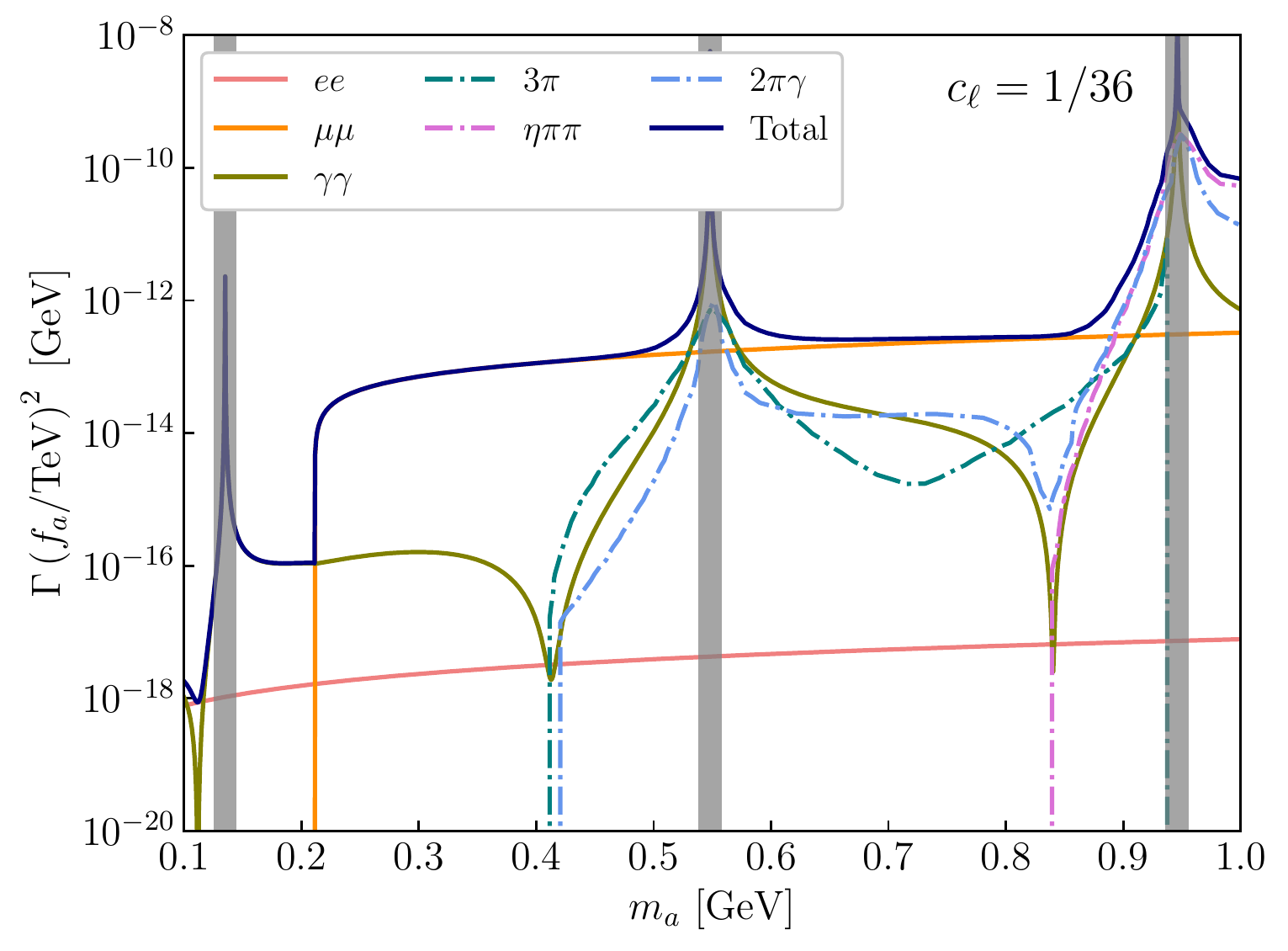}
    \includegraphics[width=0.48\textwidth]{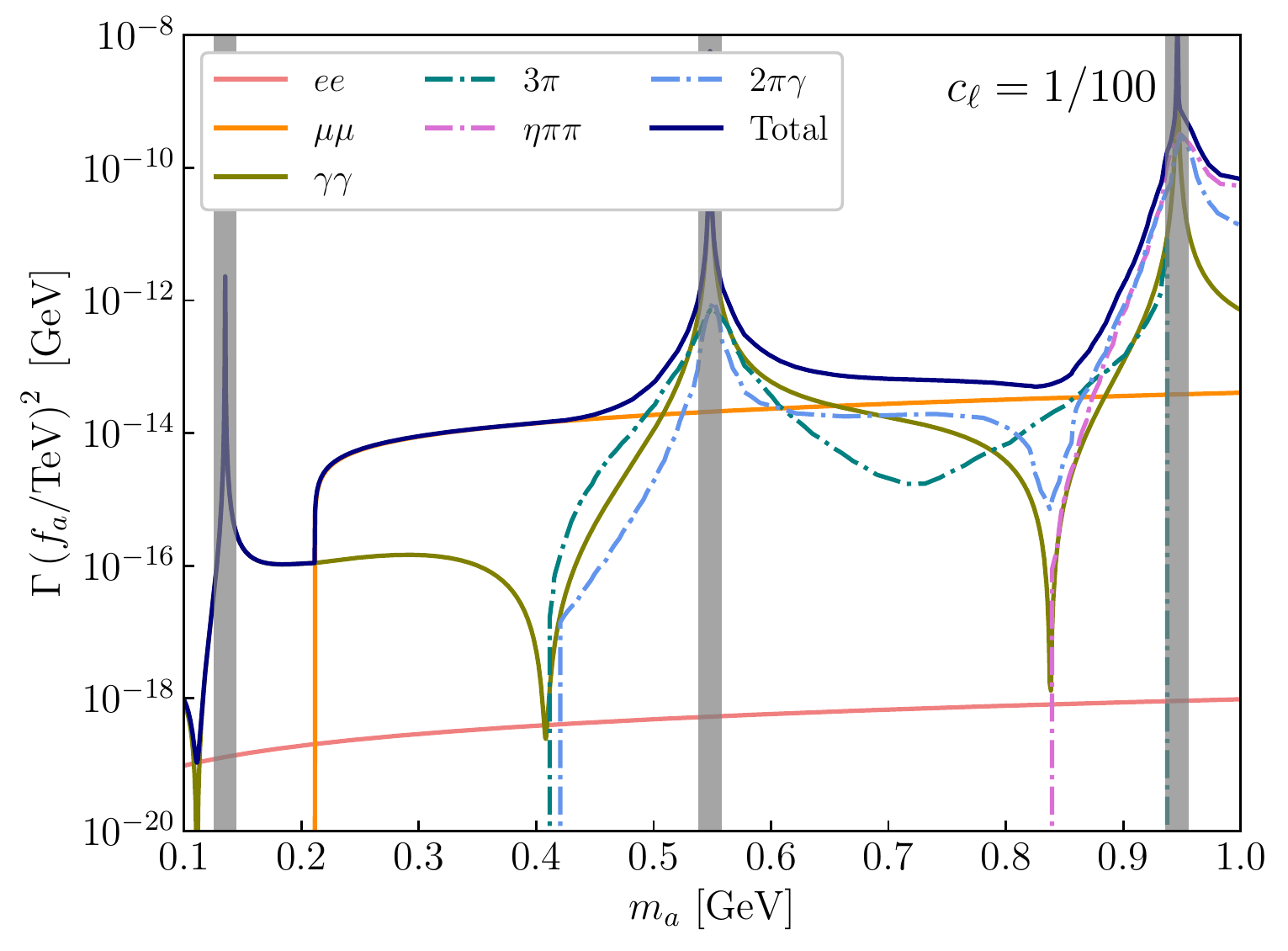}
    \caption{Contribution to axion decay width from various modes.}
    \label{fig:summary_modes}
\end{figure}

\paragraph{Summary including hadronic modes.}
In Fig.~\ref{fig:summary_modes}, we show the contribution to the axion decay modes coming from various channels, including exclusive hadronic modes: $a\rightarrow 3\pi$, $a\rightarrow\eta\pi\pi$ and $a\rightarrow\pi\pi\gamma$~\cite{Aloni:2018vki}. The relative contributions of these hadronic modes depends upon the axion-lepton coupling and we show the results for the two benchmark couplings $c_\ell=1/36$ and $c_\ell=1/100$. We can not treat the axion-SM pseudoscalar mixings perturbatively very near the SM pseudoscalar masses (denoted by gray bands), and therefore exclude these regions from our analysis.
We also see from Fig.~\ref{fig:summary_modes} that the diphoton mode does not contribute appreciably to the axion decay width unless the axion mass is very close to the SM pseudoscalar masses. These results show that the branching ratio into dimuons is significant for axions masses considered in this work.

\bibliography{supp.bib}